%% file: main.tex
\documentclass[conference]{IEEEtran}
\IEEEoverridecommandlockouts

\usepackage[hyphens]{url}
\usepackage{hyperref}
\usepackage{tabularray}
\usepackage{xcolor}
\usepackage{orcidlink}

\input{commands.tex}

\begin{document}

\title{Ritgard: T(r)opical Islands of \\ Socio-Technical Artifacts on GitHub}

\author{%
    \IEEEauthorblockN{%
        Adam Štěpánek\IEEEauthorrefmark{1}\orcidlink{0009-0008-9388-2546},\,%
        Marco Raglianti\IEEEauthorrefmark{2}\orcidlink{0000-0002-6878-5604},\,%
        Jan Byška\IEEEauthorrefmark{1}\IEEEauthorrefmark{3}\orcidlink{0000-0001-9483-7562},\,%
        Barbora Kozlíková\IEEEauthorrefmark{1}\orcidlink{0000-0003-0045-0872},\,%
        Michele Lanza\IEEEauthorrefmark{2}\orcidlink{0000-0003-4391-0197}}
    \IEEEauthorblockA{%
        \IEEEauthorrefmark{1}Visitlab, Masaryk University, Brno, Czech Republic
        \qquad
        \IEEEauthorrefmark{2}REVEAL @ Software Institute --- USI, Lugano, Switzerland}
    \IEEEauthorblockA{
        \IEEEauthorrefmark{3}VisGroup, University of Bergen, Bergen, Norway}
}

\maketitle

\begin{abstract}
A software project is more than just code. Non-code artifacts often document the human processes and decisions behind source code. The rationale behind a library change, an architectural decision, a problem encountered by a user are all examples of information typically present in socio-technical artifacts (STAs), created and persisted in channels separate from the repository itself (yet sometimes very close---\eg GitHub Issues with GitHub repositories). These STAs are a trove of information about the project's architecture and its evolution, containing details and insights that code alone cannot provide. Unfortunately, this information is not easily extracted and explored as STAs are frequently fragmented over different communication channels, and are written in natural language.

We present \ritgard, a tool that mines GitHub repositories for their STAs, namely Issues, Pull Requests, and Discussions, and visualizes them as 3D islands covered with trees. Each tree represents a single artifact and each island is a topic extracted from the artifacts through a combination of text embedding and text summarization. The terrain of the islands rises out of the ocean as the topic becomes active and sinks back in when it becomes stale, thus depicting the evolution of features and concerns throughout the project's lifetime. We describe the tool's usage and implementation, showing the numerous technical challenges behind \ritgard's visualization.

\end{abstract}

\begin{IEEEkeywords}
software visualization, topic modeling, software repository mining, socio-technical artifacts, GitHub
\end{IEEEkeywords}



\input{sections/1_introduction.tex}
\input{sections/2_related_work.tex}
\input{sections/3_tropical_islands.tex}
\input{sections/4_examples.tex}
\input{sections/5_conclusion.tex}

\bibliographystyle{IEEEtran}
\bibliography{IEEEabrv,references.bib}

\end{document}

%% file: commands.tex
\usepackage{xspace}

\newcommand{\urlttlight}[1]{\url{#1}}
\newcommand{\urltt}[1]{\textbf{\url{#1}}}

\newcommand{\ie}{\emph{i.e.,}\xspace}
\newcommand{\eg}{\emph{e.g.,}\xspace}

\newcommand{\etal}{\emph{et~al.}\xspace}

\newcommand{\ritgard}{{\sc Ritgard}\xspace}
\newcommand{\replipackage}{{\small\urltt{https://doi.org/10.6084/m9.figshare.32346822}}}

\definecolor{tablelightgray}{gray}{0.87}
\definecolor{tabledarkgray}{gray}{0.2}

\usepackage{tikz}

\newcommand\encircle[1]{%
  \tikz[baseline=(X.base)] 
    \node (X) [draw, shape=circle, inner sep=-1pt] {\strut #1};}

\usepackage{etoolbox}
\AtBeginEnvironment{thebibliography}{%
   \interlinepenalty10000%
}

\newcommand{\Description}[1]{}

\usepackage{doi}

\usepackage{csquotes}

\renewcommand{\mkbegdispquote}[2]{\itshape\openautoquote}

%% file: sections/1_introduction.tex

\vspace{-2mm}

\section{Introduction} \label{sec:introduction}

Code does not tell the whole story of a software project. Non-code artifacts, such as documentation, bug reports, code reviews, architectural decision records, and even e-mail and other messages describe the project's shape and history from another perspective, which may be  disconnected from the code repository. These artifacts form the \textit{documentation landscape}~\cite{raglianti_2023} of the project, a rugged environment scattered over different communication channels (\eg GitHub, e-mails, Slack, Discord) that are volatile and ever-changing, especially for actively developed large and long-lived projects.

GitHub is the dominant collaborative development platform~\cite{octoverse_2024} and its built-in communication channels are the backbone of the typical documentation landscape of GitHub projects. Its \textit{Issues}, \textit{Pull requests (PRs)}, and \textit{Discussions} are used to manage and document the project's development. Issues describe bugs and desirable features~\cite{github_2026a}. PRs are used to review changes to the project's code~\cite{github_2026c}. Discussions offer a forum-like environment where developers and users can meet~\cite{github_2026b}. Since these communication channels bridge the gap between society and technology, we consider each Issue, PR, and Discussion to be a \textit{socio-technical artifact} (STA)~\cite{gregor_2013,drechsler_2015}.

STAs of a project depict the social discourse that surrounds the repository, its development processes, architectural decisions, and community reception. They also describe this reality over time, thus recording the discourse and the project's evolution. This viewpoint is naturally useful in many circumstances. For example, when developers need to understand the project's purpose and the forces that shape(d) its development. Project managers need to see the project's current state and progress towards a future milestone. Users need to keep up-to-date with the project to know if it is still alive and without any security vulnerabilities. Therefore, having at least a high-level understanding of a project's STAs is useful for many stakeholders. However, getting this understanding is cumbersome, since GitHub STAs are written in natural language and, for large repositories, there simply may be far too many STAs to read without a strategy and proper tool support.

\begin{figure}[ht]
    \vspace{-3mm}
    \includegraphics[width=\linewidth]{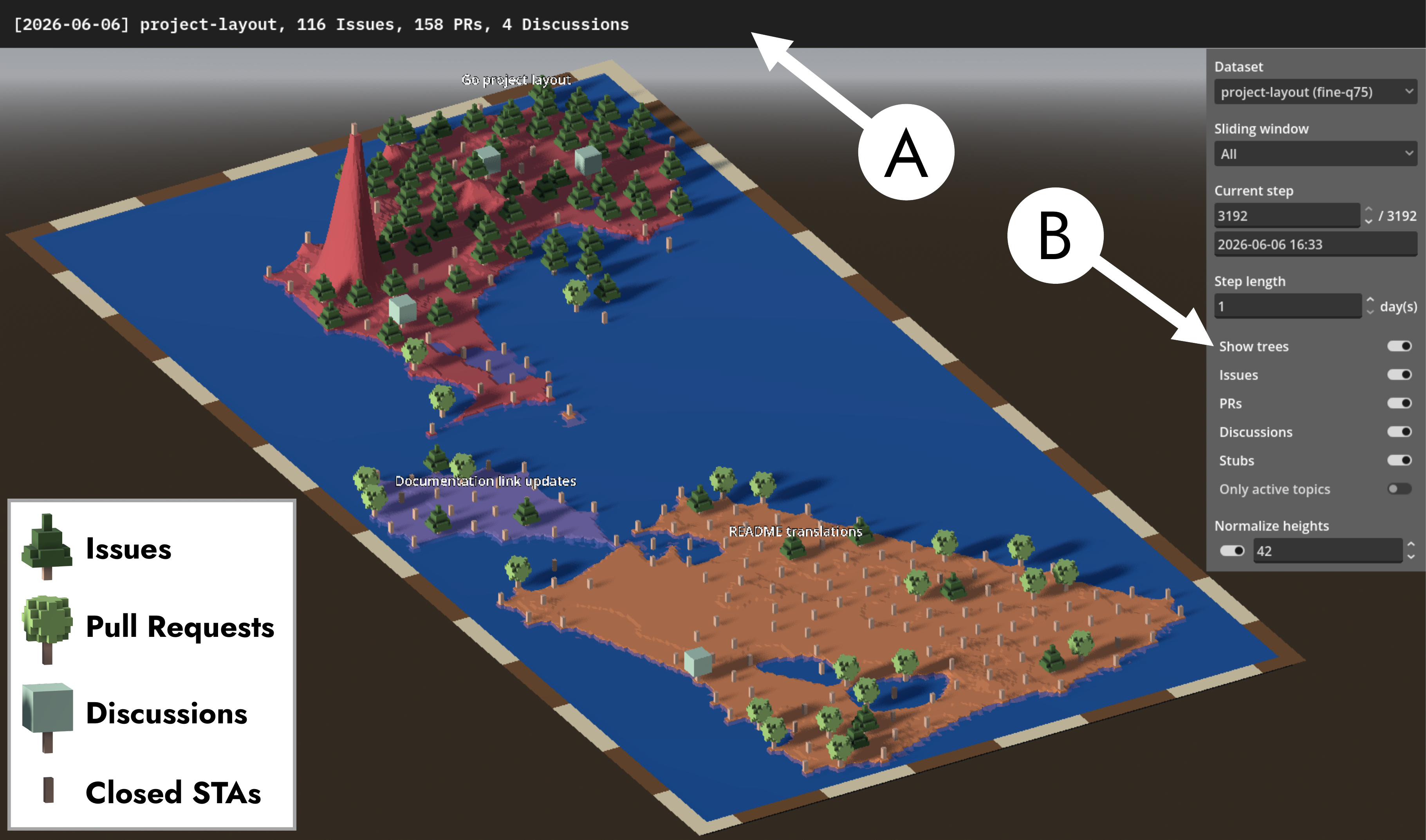}
    \caption{The \texttt{golang-standards/project-layout} repository in \ritgard. Issues, PRs, and Discussions are visualized as trees. Islands group them by topic. The UI includes a status bar (A) and a configuration side panel for the visualization parameters (B).}
    \label{fig:ui}
\end{figure}

Recent developments in text embedding and summarization using large language models (LLM) provide a practical way to solve both issues. Embedding models have proven to be capable enough to compare semantic similarity of sentences and texts~\cite{reimers_2019,mteb}. LLMs can assign topics to clusters of textual documents via their summarization capabilities~\cite{grootendorst_2022,pham_2024}. Yet, even with these advances, there remains the issue of assembling the STAs, their topics, metadata, and history in a readable, informative, high-level, and playful representation~\cite{guitard_2005,dal_sasso_2017}. In \autoref{fig:ui}, we show an example visualization and \ritgard's user interface (UI).

In our prior work~\cite{mainpaper}, we presented a visualization approach providing an overview of the STAs of a GitHub repository and evaluated its readability and usefulness in a user study. Our approach turns each topic into a 3D island in an ocean, and each STA into a tree. Here, we focus on the technical details of \ritgard---the prototype implementation of our visualization design, also used in the user study. We present its data processing pipeline from the GitHub repository to the interactive 3D visualization, provide examples, and suggest directions for future development.

%% file: sections/2_related_work.tex
 
\section{Related Work} \label{sec:related-work}

Existing tools and prior research focus on broad usage of STA types or deep analysis of individual STAs. To the best of our knowledge, there is no tool that provides a high-level overview of the three main STA types of a GitHub repository, let alone one with customized evolutionary visualizations.

Zhang \etal studied the reasons for PR acceptance or rejection~\cite{zhang_2023}. They found that the most important factor in the decision is the social distance between the author and the integrator, and that automated tools often replace the role of comments. Hata \etal conducted a study of GitHub Discussions, a newly added STA type back then, and discovered that it is essential to set proper guidelines for this platform, so indirectly for the creation of such STAs, and that core developers' participation in the discourse can make a difference~\cite{hata_2021}. Hao \etal trained a model for recommending ``good first issues'' for new contributors~\cite{hao_2022}, confirming the importance of STAs for triaging and bug fixing but also for community building and knowledge transfer. Siddiq \etal used topic modeling to assign labels to Issues, showing that these techniques are a natural fit for the NL in GitHub STAs~\cite{siddiq_2022}.

\textbf{Visualizations:} Fiechter \etal introduced \textit{issue tales}, a 2D visualization focusing on the issue lifecycle in views with various granularities~\cite{fiechter_2021}. Issue tales visualize issue metrics (\eg size, duration) and their connections, but disregard contents and topics. Kuhn \etal used machine learning methods to produce thematic maps of source code~\cite{kuhn_2008}. Although, they focused only on source code, their work inspired \ritgard.

GitHub's native UI includes a tabular view of STAs, providing useful details (\eg titles, labels, assignees). However, it is limited to a handful of items per page, textual in nature, and thus fails at providing a holistic high-level overview.

%% file: sections/3_tropical_islands.tex
 
\section{T(r)opical Islands} \label{sec:tropical-islands}

\ritgard is a visualization tool implementing the concept of \textit{t(r)opical islands}, enriched by a suite of scripts for data mining and processing. It handles everything, from mining the STAs from GitHub, to data pre-processing, topic modeling, layout and terrain generation, rendering, and user interaction. It provides a high-level overview of a snapshot of the project's discourse and facilitates interactive exploration of its evolution. \ritgard is a prototype mainly aimed at developers, but can be useful to other stakeholders, such as project managers, and even end users (\eg prospective adopters of a library assessing non-code assets, project maturity, and recurring issues).

\subsection{Visualization Design}

\ritgard visualizes the STAs of a GitHub repository as a 3D terrain consisting of tree-covered islands in a rectangular ocean (\autoref{fig:ui}). Each tree represents an individual STA, with different sources mapped on different tree types. Islands group the STAs according to their prevailing topic and the distance between trees corresponds to their semantic similarity.

\textbf{Depicting evolution:} The height of the terrain is proportional to the activity of STAs that stand on it. For example, heavily discussed Issues with hundreds of comments result in tall hills, whereas unanswered Discussions will only be small mounds of earth, almost at sea level. STA activity is measured throughout the project's entire history or is constrained to a specific period using a configurable \textit{sliding window}, which takes a certain time span (\eg a year) leading up to the currently visualized point in time. As this window slides through time, the project's evolution unveils itself in front of the user and islands rise out of ocean and sink back into it, showing which topics are active at any given time.

\textbf{Tree types:} STAs are represented using three types of 3D tree glyphs (\autoref{fig:ui}, bottom-left corner). Trees with conical tops imply an Issue, ball-top trees are PRs, and cube-top trees are Discussions. There are also stubs---trees with no treetop---that can be toggled to stand for STAs that have already been closed (either through acceptance or rejection) and thus have (unless reopened) reached the end of their life.

\textbf{Topics and outliers:} Islands group together STAs with a common topic. Each island is also assigned a random color from a preselected palette to further enforce the sense of separation among islands and topics. While focused communication is encouraged, nothing prevents the STA authors from covering multiple topics, especially when there are cross-cutting concerns. So, in instances without a prevailing topic, the STA is classified as outlier and its tree is put on a solitary voxel-based rock in the ocean to imply it standing out.

\textbf{Interactive exploration:} The user is free to roam the islands landscape with an orthographic camera and explore the STAs through mouse and keyboard interactions. When they encounter an interesting STA and desire its closer inspection, they can trigger an action that opens the artifact in their default web browser. They may also change the size of the sliding window (\ie the length of the aggregated and visualized time period) and shift it backward and forward in time, thus enabling an analysis of the project's evolution through the animation of the landscape terrain.

\begin{figure*}[ht]
    \centering
    \includegraphics[width=\textwidth]{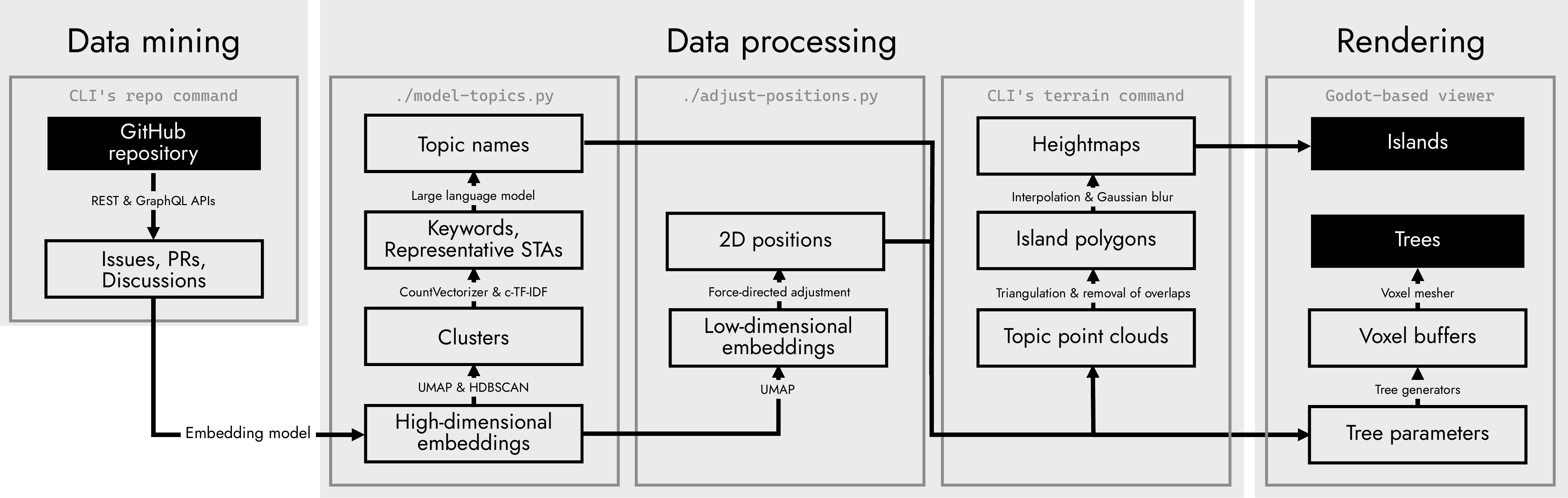}
    \caption{\ritgard's pipeline, consisting of data mining, data processing, and rendering.}
    \label{fig:pipeline}
\end{figure*}

\subsection{User Interface}

\ritgard's data mining, data processing, and terrain generation steps are implemented as separate scripts, providing a command-line interface, configurable with arguments and options. The outputs of these scripts can then be visualized in an application with a graphical user interface with two key elements: the status bar and the side panel.

The hover bar (\autoref{fig:ui}~\encircle{A}) sits at the top of the GUI and displays the name and basic information about the STA or topic that is currently being hovered over with the mouse.

If no object is hovered over, the currently visualized point in time and basic statistics about the current view are shown.

The side panel (\autoref{fig:ui}~\encircle{B}) covers all configurable aspects of the visualization. The user selects the currently visualized codebase from a list of mined and processed datasets. They may also decide to, for example, hide all trees and focus solely on the terrain, to disable showing closed STAs as tree stubs, or to normalize the height of the terrain to specific values, to either emphasize the height differences or limit occlusion.

The GUI is intentionally minimalistic, since the focus of this prototype is to evaluate the visualization design. For the interactions, it includes mouse and keyboard shortcuts, summarized in a cheatsheet available in the replication package.

\subsection{Architecture and Pipeline}

\ritgard follows a three-stage pipeline (\autoref{fig:pipeline}): data mining, data processing, and rendering. The selected GitHub repository is mined, then the extracted STAs are processed, their text is embedded, clustered, and turned into terrain data, finally, the terrain is rendered in an interactive viewer.

Each stage involves the execution of one or more scripts or programs. The data mining is performed using a C\# console application, utilizing GitHub's REST and GraphQL APIs. REST is used for Issues and PRs, whereas GraphQL for Discussions, since those are unavailable from the REST API. Due to GitHub's strict rate limiting~\cite{github_2026d}, this stage can be the pipeline's bottleneck for large repositories.

In the data processing stage, the STAs extracted from GitHub's API undergo several transformations using two Python scripts and a C\# console application. First, they are stripped of any links and Markdown syntax and embedded into high-dimensional vectors. The embedded text comprises the STA title, its labels, its main body, and/or all of its comments. Except for the STA title, all elements may be excluded to hasten the embedding step. The embedding model is also configurable, but by default the \textit{Qwen3-Embedding-8B}~\cite{zhang_2025} model is chosen due to its performance in the semantic similarity task in the Massive Text Embedding Benchmark~\cite{mteb}.

These embeddings are then reduced to low-dimensional vectors using UMAP~\cite{mcinnes_2018} and clustered with HDBSCAN*~\cite{mcinnes_2017}. These two algorithms were selected as part of the BERTopic framework~\cite{grootendorst_2022} due to their ease of use and successful application in previous research involving topic modeling~\cite{wu_2025, lezhnina_2023}. Then, an LLM is tasked with generating a short topic name for each cluster, using a prompt with the titles of the cluster's most representative STAs, keywords as given by c-TF-IDF~\cite{grootendorst_2022}, and the repository's GitHub keywords. UMAP is used again to reduce the embeddings to two dimensions, which eventually become the positions of the tree glyphs. First, however, they undergo a series of transformations to minimize tree overlap and constrain map size.

\emph{Terrain:} The third and final data processing script calculates heightmap textures for each topic-island and for each predefined sliding window length. The terrain of each island is computed using a concave Delaunay triangulation (similar to alpha shapes~\cite{edelsbrunner_1983}) of all STA positions. Triangles that overlap with other islands or with STA outliers are removed, resulting in a believable island outline. Then, each position on a grid inside a remaining triangle is interpolated using values from the triangle's corners and blurred to remove sharp edges. The interpolation and blur is done for each ``step'' of the visualization, which is the duration between two consecutive points in time. The length of this step is configurable and is set to 24 hours by default. While the use of this terrain generation script limits the sliding window to only a handful of predefined lengths, it severely lowers the computational requirements of the visualizer at runtime. This tradeoff allows the user to move through the project's history at a reasonable pace.

\begin{figure*}[t]
    \centering
    \includegraphics[width=0.95\linewidth]{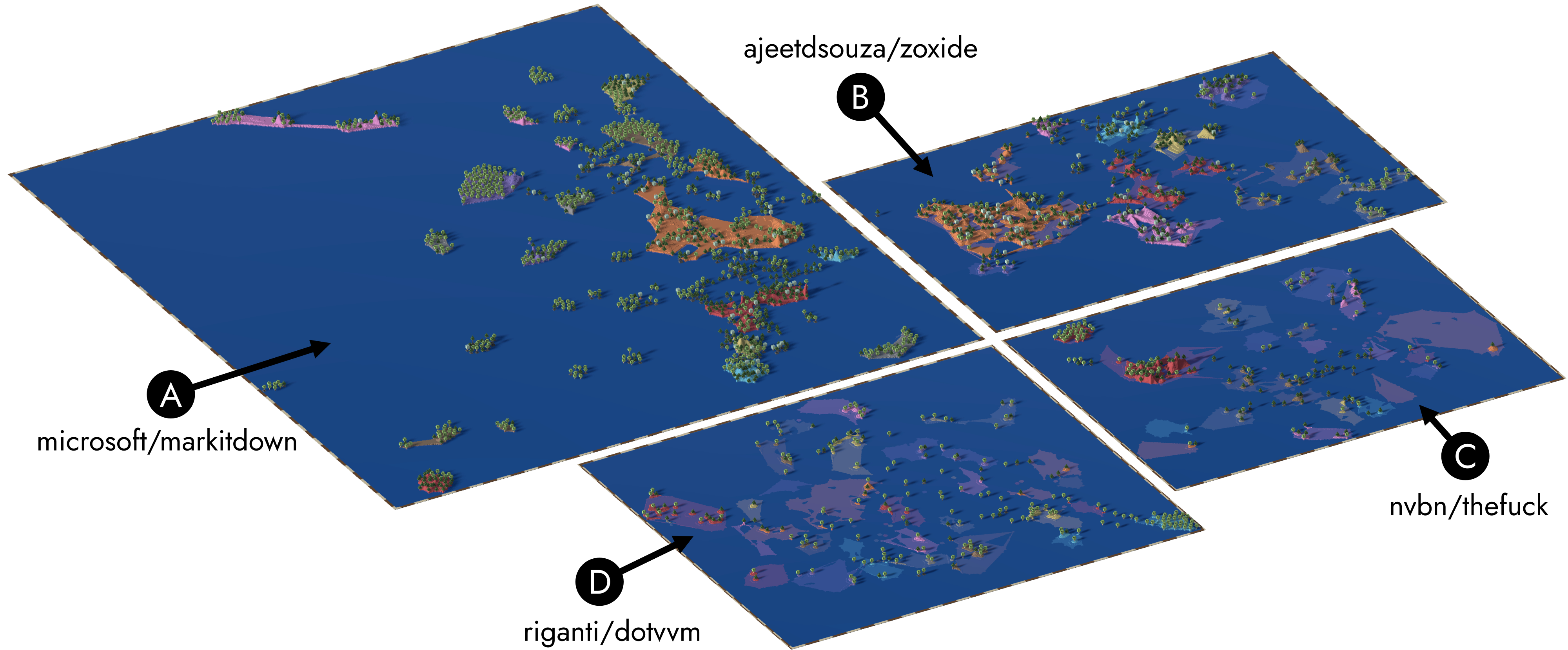}
    \caption{Examples of \ritgard's visualization, showing the last two years of active STAs in 4 software projects (A-D). Same image scale for comparison.}
    \label{fig:examples}
    \vspace{-2.5mm}
\end{figure*}

\emph{Rendering:} The final stage is the rendering. Implemented in the Godot game engine, the visualizer acts as a window on the islands of STAs mined and processed before. The visualizer itself renders the landscape and handles user interaction. It is a standalone program relying on modern rendering APIs and, therefore, can utilize the full rendering potential of a dedicated GPU, which is crucial for smooth performance and scalability of \ritgard.

The prototype's pipeline is not a one-size-fits-all solution. Its fragmented nature reflects the need for each stage to run on different hardware or with different permissions. For example, data mining requires GitHub access tokens, which are sensitive information. The topic modeling step may require high-end GPUs, depending on the embedding model, and thus runs on a different machine than the visualizer, whose requirements, in turn, include neither access to GitHub APIs nor special hardware. Therefore, \ritgard's pipeline is best ran in two or three distinct stages, so that data mining and processing runs independently of the visualizer, akin to a thin client.

%% file: sections/4_examples.tex
 
\section{Examples} \label{sec:examples}

\autoref{fig:examples} shows four examples of \ritgard's visualization. We chose the repositories based on the first author's familiarity with them and integrating with results from SEART-GHS~\cite{seart_2021}, with filters on number of STAs and stars on GitHub. They were selected to illustrate how the visualization reflects the differences between them. All four images show the same two-year period between June 2024 and June 2026 and use the same scale for comparison.

Even without the interactive pan and zoom available in the tool, these snapshots provide a few insights. For example, \encircle{A} is the most active of the four, judging by its size and tree count. However, \encircle{A} is also the youngest as there are no submerged landmasses, indicating that the project has been created in the visualized time span. Projects \encircle{B} and \encircle{C} are both command-line utilities of similar sizes (in terms of STAs), however, \encircle{B} has been considerably more active lately, with several active topic-islands above sea level. \encircle{C} and \encircle{D} share similar activity levels, but PRs (ball-top trees) are much more prevalent in \encircle{D}, indicating that its issues are likely discussed elsewhere. While \encircle{A} and \encircle{B} have adopted Discussions, there are no cube-top trees on the islands of the other projects, suggesting that either their communities are small or the developers have not enabled the feature and could consider its potential impact.

%% file: sections/5_conclusion.tex
 
\section{Conclusion \& Future work} \label{sec:conclusion}

The socio-technical artifacts of a software project tell its story from a perspective that its source code cannot convey. On GitHub, Issues, Pull Requests, and Discussions document the project's lifecycle, the key decisions of its developers, and the struggles of its users. However, these insights are difficult to access due to the sheer number STAs, their use of natural language, and the lack of existing tools to process them and provide a visually rich overview.

With  \ritgard, we have shown that a combination of topic modeling and 3D visualization techniques can produce a high-level overview of a project's STAs. \ritgard turns these artifacts into a landscape of islands representing their topics. It captures the present state of the project and its entire evolution from the perspective of its STAs. With polish and refinement, such a tool can be a practical companion to both the engineers working on the visualized project and to its users, striving to understand key non-functional implications.

We envision extending the visual metaphor to cover more STA metadata (\eg reason for STA closure) and polishing the visualizer's user experience. The tool's pipeline should also be refactored into a client-server setup, so that it can be invoked with a single button press within the visualizer itself, greatly decreasing the tool's barrier for entry of the current scripts. Finally, we want to experiment with smaller embedding models and LLMs to find the smallest models that produce good results on ``affordable'' hardware.

\textbf{Replication package:} To ensure the verifiability of our work, the tool, a demonstration video, and example datasets are available at \replipackage.

\textbf{Acknowledgements:} Computational resources were provided by the e-INFRA CZ project (ID:90254), supported by the Ministry of Education, Youth and Sports of the Czech Republic. We also gratefully acknowledge the financial support of the Swiss National Science Foundation (SNSF) through the project ``FORCE'' (SNF Project No. 232141).